\documentclass[pdflatex,sn-mathphys-num]{sn-jnl}

\usepackage{graphicx}%
\usepackage{multirow}%
\usepackage{amsmath,amssymb,amsfonts}%
\usepackage{amsthm}%
\usepackage{mathrsfs}%
\usepackage[title]{appendix}%
\usepackage{xcolor}%
\usepackage{textcomp}%
\usepackage{manyfoot}%
\usepackage{booktabs}%
\usepackage{algorithm}%
\usepackage{algorithmicx}%
\usepackage{algpseudocode}%
\usepackage{listings}%
\usepackage{graphicx}
\usepackage{svg}

\theoremstyle{thmstyleone}%

\theoremstyle{thmstyletwo}%

\theoremstyle{thmstylethree}%

\begin{document}

\author*[1]{\fnm{Joel} \sur{H\"atinen}}\email{joel.hatinen@gmail.com}
\author[1]{\fnm{Renan} \sur{Pires Loreto}}
\author[1]{\fnm{Arvind} \sur{Kumar}}
\author[1]{\fnm{Jani} \sur{Taskinen}}
\author[1,2]{\fnm{Mika} \sur{Prunnila}}

\affil[1]{\orgname{VTT Technical Research Centre of Finland Ltd.}, \orgaddress{\street{Tietotie 3}, \city{Espoo}, \postcode{02150}, \country{Finland}}}

\affil[2]{\orgdiv{presently at}, \orgname{Semiqon Oy}, \orgaddress{\street{Tietotie 3}, \city{Espoo}, \postcode{02150}, \country{Finland}}}

\title[Electronic cooling of a TiW alloy normal-metal island using Nb-based superconducting tunnel junctions] {Electronic cooling of a TiW alloy normal-metal island using Nb-based superconducting tunnel junctions}

\abstract{TiW thin films remain non-superconducting down to millikelvin temperatures and are compatible with large-scale CMOS manufacturing, making them attractive for cryogenic and quantum tunnel-junction devices, such as normal-metal--insulator--superconductor (NIS) thermal sensors and electron refrigerators. We demonstrate significant electronic cooling of a TiW-based normal-metal island and electron thermometry by using TiW-Al-AlO$_\mathrm{x}$-Nb NIS tunnel junction stacks. The NIS thermometer enables local electron temperature measurements from 0.5 to 8.32~K. NIS cooling is observed between 0.6 and 3.5~K, with maximum absolute and relative temperature reductions of 217~mK and 27\%, respectively. Calculations using tunnel-junction parameters obtained independently from current-voltage fits reproduce the measured optimum-voltage scale, which lies substantially below the ideal low-temperature prediction.}


\maketitle
\clearpage

Electronic refrigeration based on normal-metal--insulator--superconductor (NIS) tunnel junctions exploits the energy-selective tunneling of quasiparticles through the superconducting energy gap. By appropriately biasing the junction, high-energy quasiparticles are extracted from the normal metal, reducing its electronic temperature below that of the surrounding phonon bath. Since the first experimental demonstrations, NIS and SINIS junctions have become important tools for electronic cooling, thermometry, and radiation detection \cite{Nahum1994ElectronicJunction,Leivo1996EfficientJunctions,Giazotto2006OpportunitiesApplications}.

The operating temperature range of NIS refrigerators is fundamentally linked to the superconducting energy gap of the S electrode \cite{Muhonen2012Micrometre-scaleRefrigerators,Giazotto2006OpportunitiesApplications}. Consequently, extending electronic refrigeration towards higher temperatures requires superconductors with larger energy gaps than Al, which has traditionally been the material of choice due to its reproducible AlO$_\mathrm{x}$ barrier. Several studies have therefore investigated alternative superconductors with higher critical temperatures. In particular, Quaranta \textit{et al.} demonstrated electronic refrigeration using V-based tunnel junctions, extending cooling operation significantly above the range typically associated with Al-based devices \cite{Quaranta2011CoolingNanorefrigerators}.

Among elemental superconductors, Nb is particularly attractive due to its high critical temperature of approximately 9~K, large superconducting energy gap, and established role as a material platform for superconducting integrated circuits. Previous Nb-based NIS and Al-AlO$_\mathrm{x}$-Nb tunnel junction studies have demonstrated extended thermometric operation and high effective gap values, while also revealing the importance of proximity effects, barrier quality, and subgap non-idealities \cite{Nevala2012Sub-micronNb, Julin2016ApplicationsJunctions}. Related high-$T_\mathrm{c}$ NIS thermometry has also been demonstrated using NbN-based junctions, although the high junction resistance limited their suitability for electronic cooling \cite{Chaudhuri2013NiobiumMicrothermometer}. Recently, we have demonstrated efficient electronic cooling above 2~K using scalable Al--AlO$_\mathrm{x}$--Nb tunnel junction technology, highlighting the potential of Nb-based refrigeration for bridging the temperature gap between conventional electronic coolers and cryocoolers \cite{Hatinen2024EfficientJunctions}. 

For refrigeration and thermometry applications, it is advantageous for the cooled electrode to remain in the normal state throughout the operating temperature range. Such structures provide a well-defined electronic system for thermometry and electrothermal characterization, while avoiding additional superconducting transitions within the cooled region. A variety of normal metals and alloys have been employed as the normal-metal electrode in NIS devices \cite{Pekola1999NISRefrigeration,Giazotto2006OpportunitiesApplications,Muhonen2012Micrometre-scaleRefrigerators} including also macroscopic degenerately doped silicon pieces \cite{Mykkanen2020ThermionicBlocking}, but their fabrication is typically not compatible with scalable superconducting trilayer processes and, therefore, hinders their integration and scalable manufacturing. Recently, Luomahaara \textit{et al.} introduced a CMOS-compatible TiW-Al/AlO$_\mathrm{x}$ tunnel-junction technology which provides normal-state behavior in the TiW-Al electrode down to 20~mK while preserving the favorable tunnel-barrier properties of AlO$_\mathrm{x}$ \cite{Luomahaara2026ATechnology}. The technology is based on a scalable trilayer process, providing an attractive platform for integrated cooling, thermometry, and other tunnel-junction applications. The broader potential of TiW-based tunnel-junction technologies is further evidenced by a recent demonstration of scalable chip-scale refrigeration architectures capable of cooling the phonon temperature of a silicon substrate at operation temperatures below 1~K \cite{NorthropGrummanCorporation2026DemonstrationKelvin}.

In this work, we investigate electronic cooling in a TiW-Al normal-metal island (N) coupled to a Nb superconductor (S) through an AlO$_\mathrm{x}$ tunnel barrier (I), forming a NIS junction. We calibrate our NIS thermometer junction against the fridge sample holder temperature and use it to probe the electron temperature of the normal-metal island while biasing a double junction SINIS cooler in a broad operating temperature range of 0.6~K--3.5~K. Optimal cooling voltage is compared against a conventional tunnel-junction theory. We observe net electronic cooling up to an operating temperature of 3.5 K. The calculated optimum-voltage scale agrees reasonably well with experiment over the temperature range where the model predicts positive cooling power, while a discrepancy emerges at lower temperatures where cooling is nevertheless observed experimentally.

\begin{figure}[htbp]
    \centering
    \includegraphics[width=\linewidth]{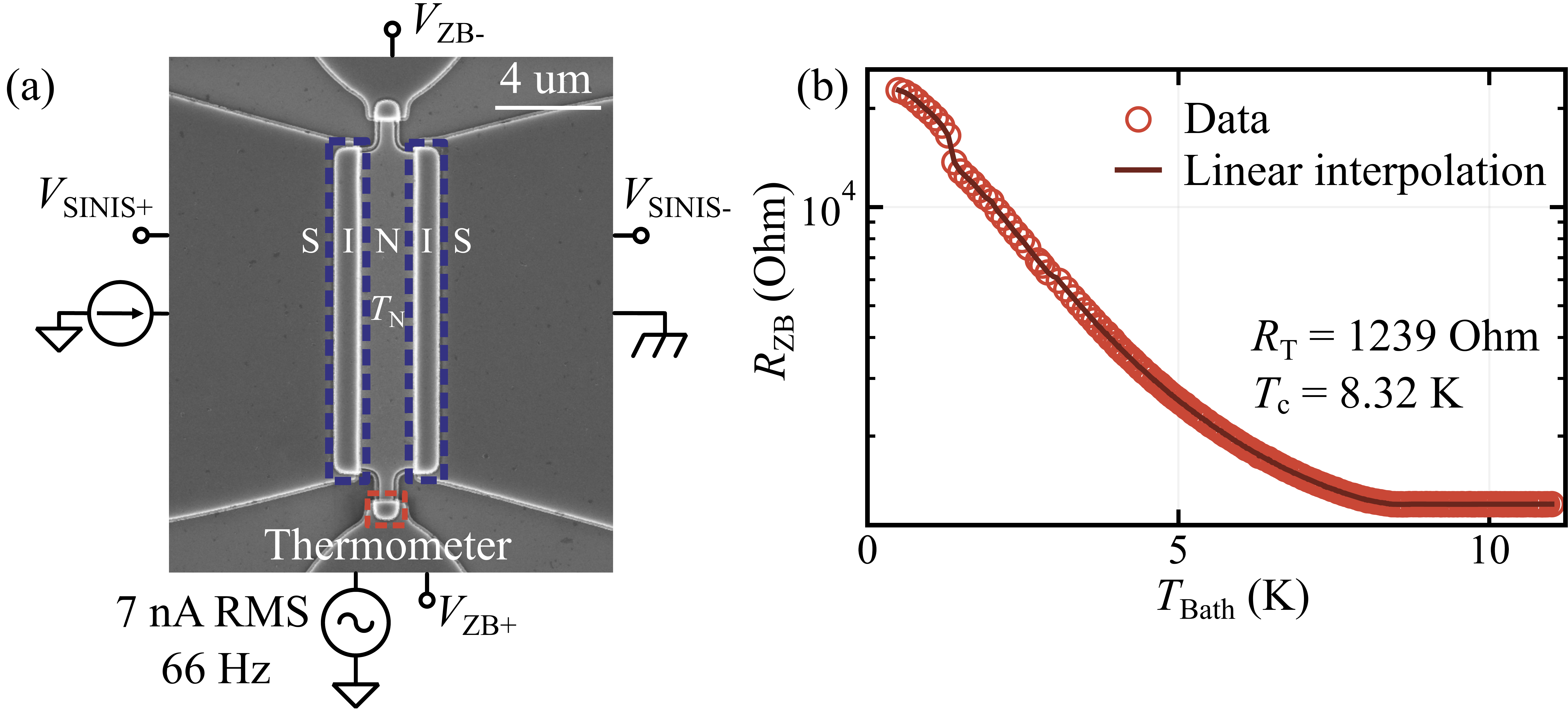}
    \caption{(a): Scanning electron micrograph of the mesoscopic cooler device with the experimental wiring. (b): Zero-bias resistance of the top junction with respect to the bath temperature. Linear interpolation between the data points is used as a thermometer calibration. }
    \label{fig:fig1}
\end{figure}

The device investigated in this work is a mesoscopic tunnel-junction cooler, shown in Fig.~\ref{fig:fig1}(a) with the experimental wiring. The structure comprises a TiW-Al base electrode connected to four Nb-Al wiring leads via AlO$_\mathrm{x}$ tunnel junctions. The larger junctions in SINIS configuration (dashed blue areas) are used for cooling due to the tunneling power scaling inversely with the resistance (see Supplementary Equation~(S2)). The small junction on the bottom of the device (red dashed area) is used for probing the electron temperature of the normal metal $T_\mathrm{N}$, while the upper junction provides the potential of the base-electrode. The device fabrication process is similar to those found in Refs.~\cite{Hatinen2024EfficientJunctions,Gronberg2017Side-wallProcess} with a few key differences to achieve functional advantages related to the thermometry and cooling operation. 

The base-electrode stack consists of a sputtered 100-nm TiW and 7-nm Al layers, with the Al surface oxidized to form the AlO$_\mathrm{x}$ tunnel barrier. The barrier is contacted by a counter electrode of 50-nm Nb and a wiring layer consisting of a 100-nm Nb and 20-nm Al. See Supplementary Section~S1 for the cross-section schematic. In contrast to the device in Ref.~\cite{Hatinen2024EfficientJunctions}, the base electrode here does not exhibit a zero-resistance state down to 300~mK (see Supplementary Section~S2). The wiring layer shows a superconducting transition at 8.63~K. 

The characteristic junction resistance was determined to be approximately 450-500~$\Omega$\textmu m$^2$ for the cooler and thermometer junctions. To probe the electron temperature in the base electrode, the zero-bias resistance $R_\mathrm{ZB}$ of the thermometer junction was characterized with respect to temperature. All low-temperature measurements were performed with a filtered 3-He adsorption cryostat, where the sample was thermalized on a base plate whose temperature $T_\mathrm{Bath}$ was measured using commercial RuOx and Cernox thermometers.

Figure \ref{fig:fig1}(b) shows the calibration of the thermometer. We observe a monotonic temperature dependence in $R_\mathrm{ZB}$ between the transition temperature $T_\mathrm{c}=$ 8.32~K, and 0.5~K. A linear interpolation of the measured resistance values is used to map the measured $R_\mathrm{ZB}$ to temperature values of the normal-metal island $T_\mathrm{N}$ when the cooler junctions are biased. The thermometer was biased with a 66~Hz sine wave at 7~nA with zero dc offset, which corresponds to a maximum of 1.1~pW of self-induced heating in the operative temperature range. The voltage drop across the thermometer junction $V_\mathrm{ZB+}-V_\mathrm{ZB-}$ was recorded using a lock-in amplifier.

\begin{figure}[htbp]
    \centering
    \includegraphics[width=\linewidth]{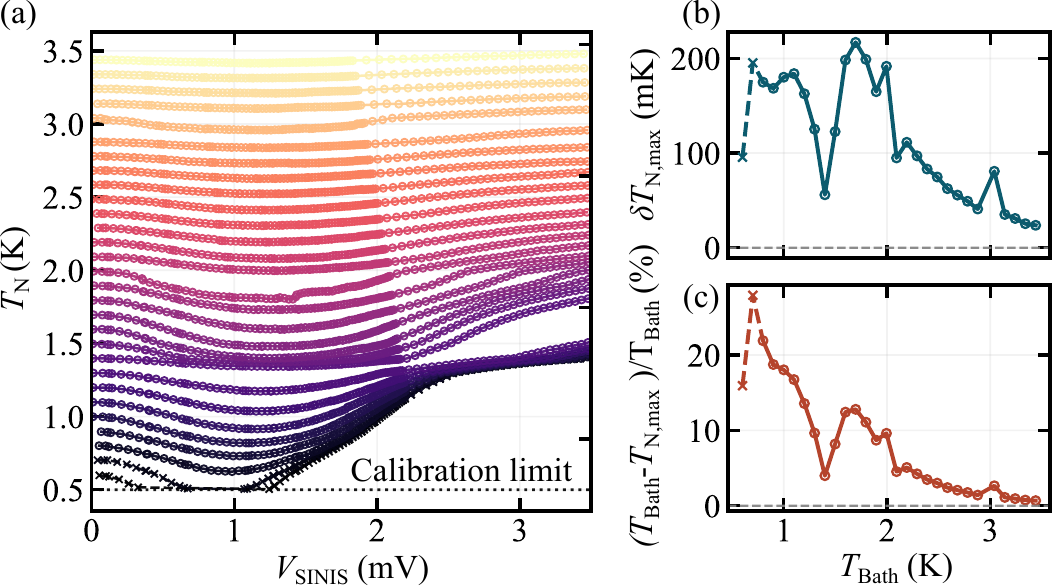}
    \caption{Cooling performance of the device. (a): Normal-metal electron temperature measured using the tunnel-junction thermometer as a function of voltage across the SINIS cooler. The operating bath temperature for each curve is recovered at zero cooler bias, i.e., $T_\mathrm{N} = T_\mathrm{Bath}$ at $V_\mathrm{SINIS}=0$.(b): Absolute cooling from the bath with a maximum of 217~mK at $T_\mathrm{Bath}=$~1.7~K. (c): Relative cooling from the bath with a maximum of 27\% at $T_\mathrm{Bath} = 0.7~\mathrm{K}$.}
    \label{fig:fig2}
\end{figure}

The cooling measurements were performed by first stabilizing the cryostat temperature $T_\mathrm{Bath}$ and then sweeping a current bias through the SINIS junction. The corresponding voltage $V_\mathrm{SINIS}=V_\mathrm{SINIS+}-V_\mathrm{SINIS-}$ was recorded in parallel using a differential preamplifier and a digital multimeter. The resulting cooling curves are shown in Fig.~\ref{fig:fig2}(a) at $T_\mathrm{Bath}=$~0.6~K to 3.5~K after translating the measured $R_\mathrm{ZB}$ values to $T_\mathrm{N}$ using the calibration from Fig~\ref{fig:fig1}(b). $T_\mathrm{Bath}$ for each curve can be observed at $T_\mathrm{N}(V_\mathrm{SINIS}=0)$. At $T_\mathrm{Bath}=$~0.6~K and 0.7~K, the cooling curves reach the lower limit of the thermometer calibration range. The corresponding data points are indicated by x-markers and dashed lines in Fig.~\ref{fig:fig2}.

The maximum cooling values at each $T_\mathrm{Bath}$ are shown in panels (b) in absolute scale and in (c) in relative scale. We observe cooling in a broad operating temperature range between 0.6~K and 3.5~K, while the maximum absolute cooling is $\delta T_\mathrm{N,max}=$~217~mK at $T_\mathrm{Bath} = 1.7~\mathrm{K}$. The maximum relative cooling is $(T_\mathrm{Bath}-T_\mathrm{N,max})/T_\mathrm{Bath} = 27\% $ at $T_\mathrm{Bath} = 0.7~\mathrm{K}$.

At $T_\mathrm{Bath} = $~1.3--1.5~K, we observe reduced cooling performance compared to the adjacent operating temperature points. Interestingly, in Fig~\ref{fig:fig1}(b), we also observe a jump in the $R_\mathrm{ZB}$ between $T_\mathrm{Bath} = 1.4~\mathrm{K}$ and 1.3~K. While we cannot assign a microscopic origin to these features, we note that the temperature of 1.4~K coincides with the critical temperature of a 20-nm-thick Al film~\cite{Hatinen2024EfficientJunctions}.

For an ideal NIS junction with a BCS density of states, the optimum cooling voltage in the low-temperature limit $T\ll\Delta/k_{\mathrm B}$ can be approximated as \cite{Muller1997ElectronApproach, Giazotto2006OpportunitiesApplications}
\begin{equation}
V_{\textrm{Opt}} \approx \frac{\Delta}{e} - 0.66\,\frac{k_{\textrm{B}} T_{\textrm{N}}}{e}.
\label{eq:vopt}
\end{equation}
For a symmetric SINIS structure containing two identical junctions in series, the corresponding total voltage is twice the single-junction value. Equation~\eqref{eq:vopt} assumes an ideal BCS density of states and does not include subgap broadening. The fitted tunnel resistances of the two cooler junctions differ by less than 2\%, whereas the fitted density-of-states parameters show greater spectral asymmetry. In particular, the fitted energy gaps are approximately 1.24 and 1.15~meV, and the corresponding Dynes parameters \cite{Dynes1978DirectSuperconductor} are approximately 0.075 and 0.16 for the left and right junctions, respectively (see Supplementary Section~S3).

\begin{figure}[htbp]
    \centering
    \includegraphics[width = .5\linewidth]{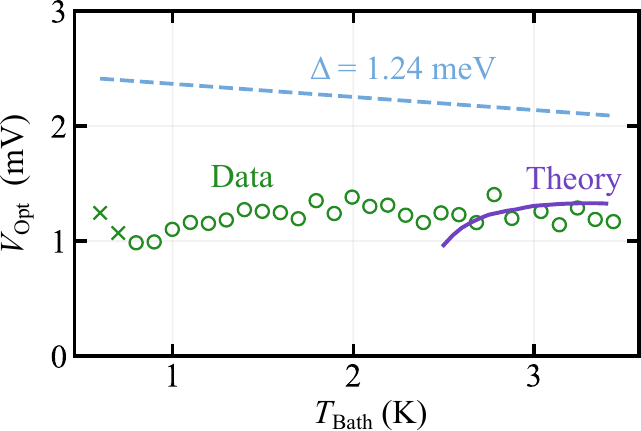}
    \caption{Optimum voltage across the SINIS cooler as a function of bath temperature. The green markers show the experimentally determined values. The blue dashed line is the ideal low-temperature approximation of Eq.~\eqref{eq:vopt} using $\Delta=1.24~\mathrm{meV}$. The purple solid line is obtained by numerically maximizing the calculated cooling power (see Supplementary Section~S4) using the separately IV-fitted tunnel resistances, energy gaps, and Dynes parameters of the two cooler junctions listed in Supplementary Table~S1. The numerical curve is shown only where the calculated maximum junction cooling power is positive. With the IV-fitted parameters, the calculation does not yield positive cooling power below approximately $2.5~\mathrm{K}$, whereas experimental cooling is observed down to $0.6~\mathrm{K}$.}
    \label{fig:fig3}
\end{figure}

Using $\Delta=1.24~\mathrm{meV}$ in Eq.~\eqref{eq:vopt} gives the dashed curve in Fig.~\ref{fig:fig3}. This ideal BCS approximation predicts a total SINIS optimum voltage between approximately 2.1 and 2.4~mV over the investigated temperature range, whereas the measured values are predominantly between approximately 1.0 and 1.4~mV. Thus, the ideal gap-based approximation does not describe the observed optimum voltage.

To account for the nonideal density of states observed in the current--voltage characteristics, we additionally calculated the junction cooling powers numerically (see Supplementary Section~S4) using the parameters obtained independently from the individual-junction IV fits in Supplementary Section~S3. For junction $i$, the normalized superconducting density of states was represented using the phenomenological Dynes density of states \cite{Dynes1978DirectSuperconductor}:
\begin{equation}
n_{\mathrm{S},i}(E) = \left| \operatorname{Re} \left[ \frac{E+i\gamma_i\Delta_i} {\sqrt{(E+i\gamma_i\Delta_i)^2-\Delta_i^2}} \right] \right|,
\label{eq:dynes_dos}
\end{equation}
where $\Delta_i$ and $\gamma_i$ are the fitted energy gap and Dynes parameter, respectively. The left- and right-junction cooling powers were evaluated separately, including their fitted tunnel resistances and density-of-states parameters, and the total SINIS optimum voltage was determined numerically under the experimental biasing condition.

The resulting calculation is shown by the solid curve in Fig.~\ref{fig:fig3}. In the temperature interval where the model yields a physically relevant cooling optimum, the calculation reproduces the measured optimum-voltage scale and its comparatively weak temperature dependence. The low measured $V_{\mathrm{Opt}}$ is therefore consistent with the junction-dependent, effectively broadened density of states obtained from the IV fits. The agreement is nevertheless incomplete. However, the same IV-derived parameter set does not yield positive junction cooling power below approximately $2.5~\mathrm{K}$, whereas cooling is observed experimentally down to $0.6~\mathrm{K}$. Agreement in $V_{\mathrm{Opt}}$ therefore shows that the density-of-states parameters capture the voltage scale of the tunneling process, but it does not establish a complete temperature-dependent description of the cooling power.

The comparison in Fig.~\ref{fig:fig3} highlights two distinct aspects of the device behavior. First, the ideal low-temperature approximation based only on the superconducting energy gap substantially overestimates the measured optimum SINIS voltage. Second, a numerical calculation using the independently IV-fitted parameters of the two cooler junctions reproduces the measured optimum-voltage scale where the model yields a cooling optimum. The low $V_{\mathrm{Opt}}$ therefore does not imply an anomalously small superconducting energy gap. Instead, it is consistent with the strong subgap broadening evident in the individual-junction IV characteristics. 

However, with the IV-fitted Dynes parameters, the maximum calculated junction cooling power becomes non-positive below approximately $T_{\mathrm{Bath}}=2.5~\mathrm{K}$, whereas electronic cooling is observed experimentally down to $0.6~\mathrm{K}$. One possible origin of the discrepancy is that the effective parameters extracted from the IV characteristics at $T_{\mathrm{Bath}}=1.5~\mathrm{K}$ are not temperature-independent microscopic properties of the junctions. In particular, a single Dynes parameter may represent the combined influence of intrinsic subgap states, proximity-modified spectral features, environmental broadening, and other nonideal transport processes \cite{Pekola2010Environment-assistedStates, Herman2016MicroscopicStates}. Applying the fitted values unchanged over the full 0.6--3.5~K cooling range is therefore a modeling assumption rather than an experimentally verified temperature dependence. Alternatively, the discrepancy may arise from processes outside the single-particle density-of-states model, including nonequilibrium quasiparticle distributions in the Nb--Al electrode stack \cite{Vasenko2009NonequilibriumJunctions, Kauppila2013Non-equilibriumJunctions}.

Non-ideal transport in Nb-based AlO$_\mathrm{x}$ tunnel junctions has been reported previously. In submicron Al-AlO$_\mathrm{x}$-Nb junctions, Julin and Maasilta observed excess subgap current that could not be described by standard single-particle tunneling theory, Dynes broadening, overheating, pair-breaking, or higher-order Andreev processes, and suggested that barrier or Nb/AlO$_\mathrm{x}$ interface states may play a role \cite{Julin2016ApplicationsJunctions}. These earlier observations support treating the large fitted Dynes parameters as effective descriptions of nonideal transport rather than as direct measures of intrinsic lifetime broadening. However, the microscopic mechanism proposed for those junctions cannot be assigned to the present device based on the available measurements.

The second notable observation is the discontinuity in the thermometer calibration occurring between 1.3~K and 1.4~K. Interestingly, this temperature coincides with the superconducting transition temperature previously observed for 20-nm-thick Al films fabricated using the same deposition method. While the coincidence suggests that the superconductivity in the Al layer may influence the local density of states and therefore the thermometer response, the present data do not establish a direct connection.

We have demonstrated electronic cooling using a Nb-Al bilayer as the superconducting electrode and a TiW-Al bilayer as the normal-metal electrode. The monotonic zero-bias resistance dependence on temperature in a broad temperature range from 8.32~K down to 0.5~K enabled low-power thermometry. The cooling was observed in an operative temperature range of 0.6--3.5~K with a single device. We measured maximum absolute and relative temperature reductions of 217~mK and 27\%, respectively.

The measured optimum SINIS voltage, where minimum electron temperature of the normal metal is reached, is substantially lower than the ideal low-temperature prediction based only on the superconducting energy gap of Nb. However, numerical calculations using the separately IV-fitted parameters of the two cooler junctions reproduce the measured optimum-voltage scale where the model yields a cooling optimum. The reduced voltage is therefore consistent with the junction-dependent density-of-states parameters, yet the same parameter set does not reproduce the full temperature range over which cooling is observed.

\backmatter

\section*{Declarations}

\subsection*{Funding}
The research was funded by the European Union’s Horizon RIA, EIC and ECSEL programmes under grant agreements No. 101113086 SoCool, No. 101007322 MatQu, and No. 101113983 Qu-Pilot. We also acknowledge financial support of Chips JU project Arctic No. 101139908 and Research Council of Finland through QMAT Centre of Excellence project No. 374172.

\subsection*{Acknowledgements}
The authors thank Antti Kemppinen, Alberto Ronzani, Pauli Virtanen, and Ilari Maasilta for useful discussions. The sample fabrication was done in VTT and OtaNano Micronova cleanroom facilities with the aid of Manika Maharjan, Annukka Alanen and Sari Ahlfors.

\subsection*{Conflict of Interest}
The authors have no conflicts of interest to disclose.

\subsection*{Author Contributions}
J.H performed experiments, data-analysis, and wrote the manuscript with input from all authors. R.P.L fabricated the device. A.K contributed to the data-analysis. J.T contributed to the device design. M.P supervised the work.

\subsection*{Data availability}
The data that support the findings of this study are available from the corresponding author upon
reasonable request.

\bibliography{references}

\renewcommand{\thesection}{S\arabic{section}} \renewcommand{\theequation}{S\arabic{equation}} \renewcommand{\thefigure}{S\arabic{figure}} \renewcommand{\thetable}{S\arabic{table}}
\clearpage
\section*{Supplementary material}

\section{Device cross-section}
\label{sect:supplementary_device}
\begin{figure}[h!]
    \centering
    \includegraphics[width=0.6\linewidth]{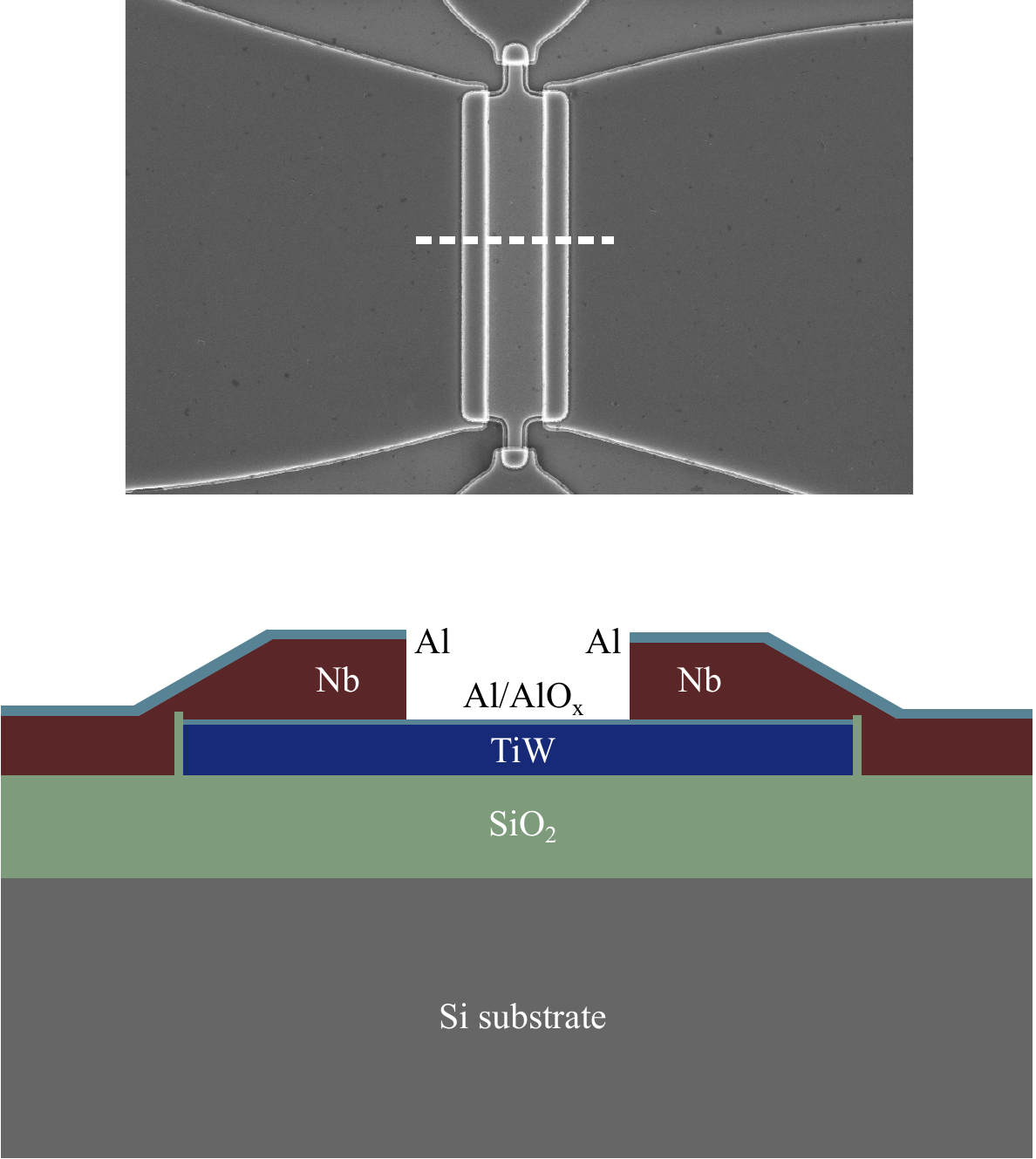}
    \caption{Cross-section illustration of the cooler device along the dashed line in the SEM image. Junction contacts are located between TiW-Al and Nb. Sidewalls of the junctions are passivated by the SiO$_\mathrm{2}$-spacers.}
    \label{fig:sfig1}
\end{figure}

The cross-section illustration of the device fabricated using a sidewall passivated spacer process \cite{Gronberg2017Side-wallProcess,Hatinen2024EfficientJunctions} is depicted on Fig.~\ref{fig:sfig1}. Thermally grown SiO$_\mathrm{2}$ is used to electrically isolate the device from the Si substrate, and SiO$_\mathrm{2}$ spacers isolate the junction sidewalls from the Nb wiring layers. Al on top of the Nb was designed similarly to Ref.~\cite{Hatinen2024EfficientJunctions} to reduce the backflow of quasiparticles from Nb back to the TiW-Al bilayer \cite{Pekola2000TrappingSuperconductor, Nguyen2013TrappingCooler}.

\section{Sheet resistances of TiW-Al and Nb-Al bilayers}
\label{sect:supplementary_sheetR}
\begin{figure}[h!]
    \centering
    \includegraphics[width=0.6\linewidth]{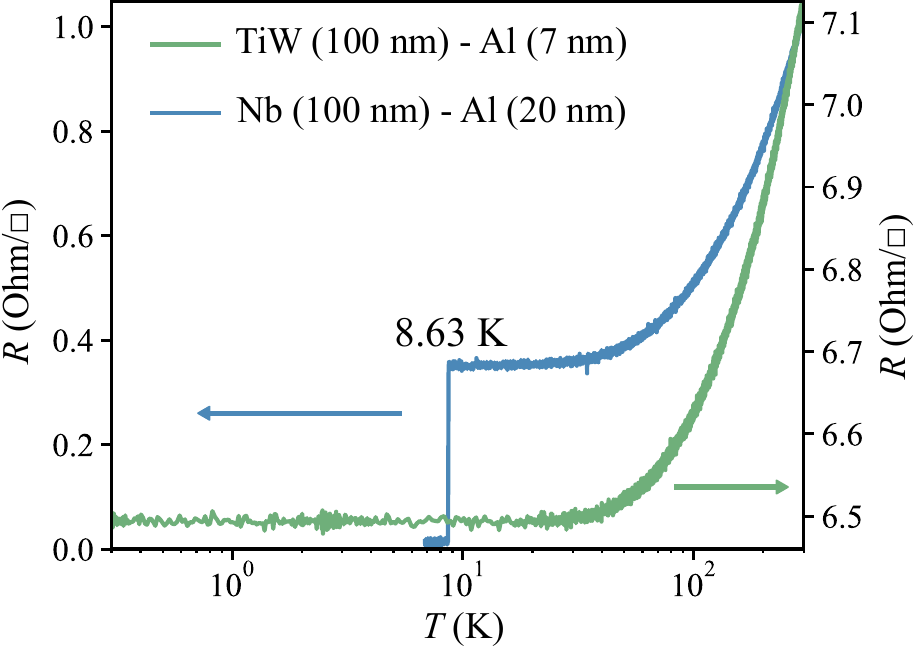}
    \caption{Sheet resistances of the TiW-Al and the Nb-Al bilayers measured from Van der Pauw structures. The TiW-based bilayer is observed to stay resistive down to 0.3~K, while the Nb-Al bilayer turns superconducting at 8.63~K.}
    \label{fig:sfig2}
\end{figure}

The sheet resistances for the TiW (100~nm) - Al (7~nm) base electrode and Nb (150~nm) - Al (20~nm) wiring layer are shown in Fig~\ref{fig:sfig2} from room temperature down to 0.3~K for TiW and 7~K for the wiring layer. The wiring layer is observed to turn superconducting at 8.63~K. The base electrode remains resistive down to 0.3~K with approx. 6.4~$\Omega$ per square. The data for both bilayers was measured in a Van der Pauw configuration using four-probe sensing and a lock-in amplifier.

\section{Cooler junction current-voltage characteristics}
\label{sect:supplementary_IVs}
The left and right cooler junctions were measured independently at a bath temperature of 1.5~K and the data was fitted using the single-particle tunneling expression \cite{Nahum1994ElectronicJunction, Leivo1996EfficientJunctions,Giazotto2006OpportunitiesApplications}:
\begin{equation}
I(V) = \frac{1}{eR_{\mathrm{T}}} \int_{-\infty}^{\infty} n_{\mathrm{S}}(E) \left[ f_{\mathrm{N}}(E-eV,T_{\mathrm{N}}) - f_{\mathrm{S}}(E,T_{\mathrm{S}}) \right] \,dE, 
\label{eq:nis_current}
\end{equation}
where $R_{\mathrm{T}}$ is the tunnel resistance, $f_{\mathrm{N}}$ and $f_{\mathrm{S}}$ are the Fermi--Dirac distributions in the normal-metal and superconducting electrodes, respectively, and $V$ is the voltage across the individual junction. The normalized superconducting density of states was represented by Dynes density of states given in Eq.~(2) of the main text.

\begin{figure}[h!]
    \centering
    \includegraphics[width=.9\linewidth]{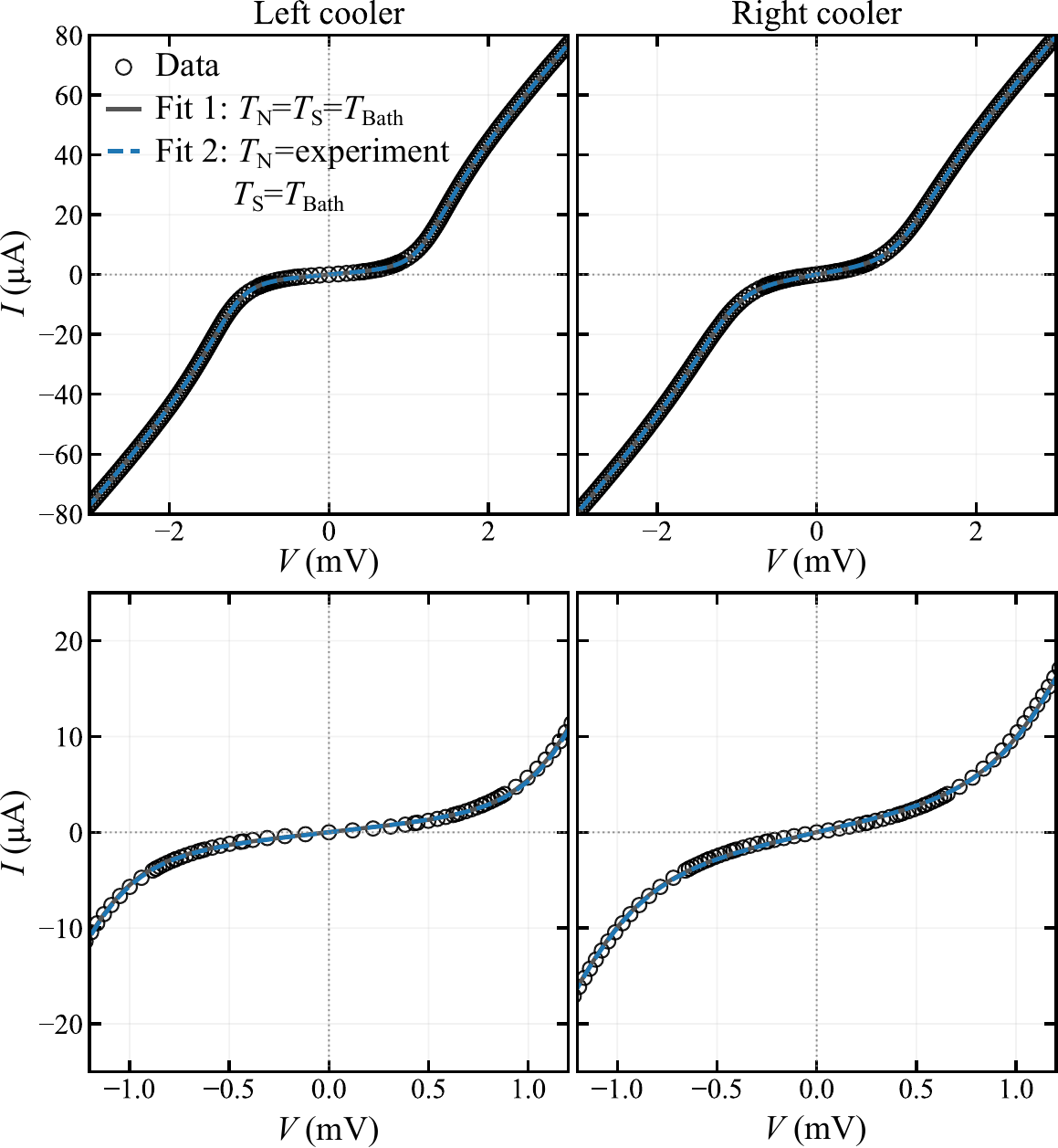}
    \caption{Current-voltage characteristics of the left and right cooler junctions at $T_\mathrm{Bath}=$1.5~K. Fit 1 corresponds to $T_\mathrm{N}=T_\mathrm{S}=T_\mathrm{Bath}$. Fit 2 uses the experimental $T_\mathrm{N}$ data.}
    \label{fig:sfig3}
\end{figure}

Full IVs and the subgap regions are shown in Fig.~\ref{fig:sfig3} together with two fit lines. Fit 1 assumes thermal equilibrium, i.e., $T_\mathrm{N} = T_\mathrm{S} = T_\mathrm{Bath}$. Fit 2 uses the experimental $T_\mathrm{N}$ data, but fixes $T_\mathrm{S} = T_\mathrm{Bath}$. The fitted tunnel resistances of the two junctions are similar, with a difference below 2\%. The fitted superconductor parameters are less symmetric: the energy gaps differ by approximately 7\%, while the Dynes parameter of the right junction is approximately twice that of the left junction. The device is therefore nearly symmetric in normal-state tunnel resistance but asymmetric within the Dynes description. The fitted parameters are summarized in Supplementary Table~S1. Additionally, a third fit was conducted by setting $T_\mathrm{S}$ as a free parameter at each bias and restricting its values to above bath temperature, i.e., $T_\mathrm{S} \geq T_\mathrm{Bath}$, but this fit effectively reduced to the fit 2.

\begin{table}[h!] \centering \caption{Fit parameters extracted from the individual cooler-junction current-voltage characteristics. Fit 1 assumes $T_\mathrm{N}=T_\mathrm{S}=T_\mathrm{Bath}$, while Fit 2 uses the experimentally measured $T_\mathrm{N}$ with $T_\mathrm{S}=T_\mathrm{Bath}$.} \label{tab:iv_fit_parameters} \begin{tabular}{llccc} \hline Junction & Fit & $R_\mathrm{T}$ ($\Omega$) & $\Delta$ (meV) & $\gamma$ \\ \hline Left cooler & Fit 1 & 35.47 & 1.235 & $7.50\times10^{-2}$ \\ & Fit 2 & 35.46 & 1.236 & $7.82\times10^{-2}$ \\ \hline Right cooler & Fit 1 & 34.82 & 1.145 & $1.57\times10^{-1}$ \\ & Fit 2 & 34.81 & 1.146 & $1.60\times10^{-1}$ \\ \hline \end{tabular} \end{table}

\section{Numerical calculation of the optimum cooling voltage}
\label{sect:supplementary_numerics}
The heat current out of the normal-metal electrode through junction $i$ was calculated using the standard single-particle tunneling expression \cite{Nahum1994ElectronicJunction,Leivo1996EfficientJunctions,Giazotto2006OpportunitiesApplications}:
\begin{equation}
\dot Q_{\mathrm{N},i} = \frac{1}{e^2R_{\mathrm{T},i}} \int_{-\infty}^{\infty} (E-eV_i)n_{\mathrm{S},i}(E) \left[ f_{\mathrm{N}}(E-eV_i,T_{\mathrm{N}}) - f_{\mathrm{S}}(E,T_{\mathrm{S}}) \right]dE,
\label{eq:Qdot}
\end{equation}
where $n_{\mathrm{S},i}(E)$ is given by Eq.~(2) of the main text. The tunnel resistances, energy gaps, and Dynes parameters were fixed to the values extracted from the individual-junction IV characteristics in Supplementary Table~S1. For each bath temperature and applied bias, the voltage division between the two junctions was determined consistently with the common series current. The net junction heat current out of the normal-metal island was obtained by summing the heat currents through the two junctions, and the theoretical $V_{\mathrm{Opt}}$ was identified from the maximum of the calculated cooling response. The electrode temperatures were set to $T_{\mathrm{Bath}}=T_\mathrm{N}=T_\mathrm{S}$, and the IV-derived parameters obtained at 1.5~K were treated as temperature independent.

\end{document}